\documentclass[12pt]{article}
\pdfoutput=1
\usepackage{textcomp}
\usepackage{color}
\usepackage{putex}
\usepackage{autobreak}
\usepackage{indentfirst}
\usepackage{graphicx}
\usepackage{float}
\graphicspath{{plots/}}
\usepackage{tabularx}
\usepackage{caption}
\usepackage{amsmath}
\numberwithin{equation}{section}
\usepackage{cancel}
\usepackage{array}
\usepackage{subcaption}
\usepackage{epstopdf}
\usepackage{enumerate}
\usepackage{cite}
\usepackage{tensor}
\usepackage{CJKutf8}
\usepackage[utf8]{inputenc}
\usepackage{rotating}
\usepackage{multirow}
\usepackage[
      colorlinks=true,
      linkcolor=red,
      urlcolor=red,
      filecolor=black,
      citecolor=red,
      ]{hyperref}
\usepackage[noabbrev]{cleveref}
\begin{document}
\newcommand{\kr}[1]{  \textbf{\textcolor{red}{(#1 --kr)}}}
\newcommand{\xz}[1]{  \textbf{\textcolor{blue}{(#1 --xz)}}}
\institution{kits}{Kavli Institute for Theoretical Sciences, University of Chinese Academy of Sciences, \cr Beijing 100190, China.}
\institution{schoolphys}{School of Physical Sciences, University of Chinese Academy of Sciences, No.19A Yuquan Road, \cr Beijing 100049, China.}
\title{More on spin-2 operators in holographic quantum mechanics}
\authors{
Shuo Zhang\worksat{\kits}${}^{,}$\worksat{\schoolphys}${}^{,}$\footnote{{\hypersetup{urlcolor=black}\href{mailto:zhangshuo163@mails.ucas.ac.cn
}{zhangshuo163@mails.ucas.ac.cn}}}
}
\abstract{We study the spectrum, unitarity bound and holographic central charge of spin-2 operators in a warped $\text{AdS}_2 \times \text{S}^2 \times \text{T}^4 \times \mathcal{I}_\psi \times \mathcal{I}_\rho$ background in the Type IIB theory. We were able to identify a class of solutions that is completely independent of the functions that define the background solution. We comment on the relation of our results to the previous ones in the literature.}
\date{\today}
\maketitle
{
\hypersetup{linkcolor=black}
\tableofcontents
}
\newpage
\section{Introduction}\label{sec: intro}

The AdS/CFT correspondence \cite{Maldacena:1997re,Witten:1998qj,Gubser:1998bc} relates string theories defined in an $\text{AdS}_{d+1}$ spacetime to superconformal field theories in $d$-dimensions. Since its inception there has been a flurry of activities in the systematic study of conformal and/or supersymmetric field theories using both field theory methods as well as supergravity techniques. Tremendous efforts have been put towards the classification of supersymmetric $\text{AdS}_{d+1}$ backgrounds in (massive) type IIA/B and M-theory. In the special case that supersymmetry is half-maximal and $d \geq 2$ there has been significant progress in the supergravity classification and the mapping of these solutions to specific quantum field theories \cite{Lin:2004nb,Gaiotto:2009gz,Ferrara:1998gv,Brandhuber:1999np,Apruzzi:2013yva,Assel:2011xz,Lozano:2019emq,Nunez:2019gbg,Legramandi:2021uds,Akhond:2021ffz}.

The AdS$_2$/CFT$_1$ case is of particular interest both for the microscopical description and understanding of extremal black holes since it appears as the near-horizon limit \cite{Bertotti:1959pf,Robinson:1959ev}, and also in the context of studying holographic defects. Despite its interesting status much less is known due to subtle points that arise in AdS$_2$/CFT$_1$ and are not present in the higher-dimensional examples \cite{Almheiri:2014cka,Denef:2007yt,Maldacena:1998uz,Harlow:2018tqv}. To begin with, when one considers the backreaction effect of finite-energy excitations, one ends up with a divergence at one of the asymptotic boundaries. In addition to that, global $\text{AdS}_2$ has two boundaries that are disconnected. Therefore, naively, it appears that the proper holographic description is in terms of two copies of the same 1-dimensional field theory. This, however, is in contradiction with the calculations of black hole entropy in string theory. Furthermore, gravity in AdS$_2$ is non-dynamical\footnote{This argument is true in the case of pure $\text{AdS}_2$. More specifically, adding matter and/or any other warp factors bypasses this hindrance, see e.g. \cite{Lozano:2020txg}.}. 

An interesting case study was presented in \cite{Lunin:2015hma}. The author presented a general class of regular geometries that asymptote to $\text{AdS}_2 \times \text{S}^2$. These are solutions that resemble wormholes and the presence of the two boundaries appears as a feature of the construction instead of a problematic description. Expanding these regular solutions results to metric perturbations very close to the ones we will consider in this work, although in a completely different string theory setup.

Recently we have a lot and exciting examples of new supergravity solutions with an $\text{AdS}_2$ factor \cite{Lozano:2020txg,Lozano:2020sae,Lozano:2021rmk,Ramirez:2021tkd,Lozano:2021fkk,Lozano:2022vsv,Lozano:2022swp,Dibitetto:2019nyz,Hong:2019wyi,Legramandi:2023fjr}. From these recent works, it was understood that the additional fluxes and warp factors lead to an $\text{AdS}_2$ theory where gravity is no longer non-dynamical. These solutions are generalizations of previous studies with constant fluxes \cite{Strominger:1998yg,Balasubramanian:2003kq,Hartman:2008dq,Alishahiha:2008tv,Balasubramanian:2009bg}. In some of the recently derived $\text{AdS}_2$ solutions, the dual field theories have been identified and several checks have been performed \cite{Lozano:2020txg,Lozano:2020sae,Lozano:2021rmk,Ramirez:2021tkd,Lozano:2021fkk,Lozano:2022vsv,Lozano:2022swp,Dibitetto:2019nyz}.  

Based on the principles of holography the spectrum of field operators is derived by considering the linearized fluctuations of the fields in the bulk for a specific supergravity solution. However, computing the complete set of linearized fluctuations for the full supergravity background is a formidable task. To this day, we have a handful of explicit examples at our disposal \cite{PhysRevD.32.389,Deger:1998nm,Eberhardt:2017fsi}.

In spite of this complication, the work of \cite{Bachas:2011xa} provided a proof that if we focus only on the spin-2 operators and their spectrum the situation is dramatically simplified. In particular, the authors proved that the spin-2 fluctuations, which result as perturbations of the Minkowski part of the geometry, satisfy an equation depending solely on the geometrical data of the background supergravity configuration. While the authors of that paper focused on an $\text{AdS}_4$ solution, a generalization of their logic to arbitrary space-time dimensions is straightforward. This laid the foundations for many
new and exciting results on the spectrum of holographic spin-2 operators across all dimensions \cite{Rigatos:2022ktp,Speziali:2019uzn,Lima:2022hji,Klebanov:2009kp,Itsios:2019yzp,Chen:2019ydk,Gutperle:2018wuk,Passias:2018swc,Passias:2016fkm,Roychowdhury:2023lxk,Apruzzi:2019ecr,Apruzzi:2021nle}. 

It is worthwhile mentioning that for theories with 8 Poincar\'e supersymmetries in AdS$_{4,5,6,7}$, the spin-$2$ fluctuations have been put into a universal treatment \cite{Lima:2023ggy}. We note, however, that only dynamics were described, and the equations were not solved except for the case of the AdS$_7$ backgrounds.

This paper is organized as the following: In section \ref{sec: prems} we briefly show the background we are going to focus on and it was derived in \cite{Lozano:2020txg}. In section \ref{sec: spin2fluctuations} we study the linear fluctuations and solve the equation of motion. We find that it becomes a Sturm-Liouville problem. In section \ref{sec: unitarityandsolutions} we calculate the unitarity bound and so-called ``minimal universal class of solutions''. Here ``universal'' means that the solutions do not depend on the warp factors of the backgrounds and ``minima'' is in the sense that we get these solutions under the condition that we saturate the unitarity bound. In the section \ref{sec: centralcharge} we compute the holographic central charge. We show our conclusions and outlooks in section \ref{sec: outlook}.
\section{Preliminaries}\label{sec: prems}
The purpose of this section is to present and discuss the supergravity backgrounds that we will examine in this work and their holographic field theory descriptions. In doing so, we also 
set up our conventions. These $\text{AdS}_2$ solutions were obtained in the work \cite{Lozano:2020txg}, which we will closely follow in the forthcoming presentation.
\subsection{The supergravity solutions}
    Let us begin by reviewing the main characteristics of the $\text{AdS}_2$ solutions that will be the focus of this work. We will be following \cite{Lozano:2020txg}, as we have already mentioned, in our presentation where this class of backgrounds was originally developed. The family of supergravity backgrounds has the schematic form $\text{AdS}_2 \times \text{S}^2 \times \text{CY}_2 \times \mathcal{I}_{\psi} \times \mathcal{I}_{\rho}$. For the purposes of our analysis we make a concrete choice for the Calabi-Yau $2$-fold and we choose it to be the $4$-dimensional torus, which means $\text{CY}_2 \equiv \text{T}^4$. This solution was obtained after a T-duality transformation in an $\text{AdS}$-coordinate from the mother massive type IIA backgrounds that assume the form $\text{AdS}_3 \times \text{S}^2 \times \text{T}^4 \times \mathcal{I}_{\rho}$ and were originally developed in \cite{Lozano:2019emq}. In the $\text{AdS}_2$ picture, $\psi$ is the T-dual coordinate and has a $\text{U}(1)$ isometry. We note that the interval $\mathcal{I}_{\rho}$ is, also, bounded as required for the $8$-dimensional manifold to be compact \cite{Lozano:2020txg}. The NS-NS sector of the $\text{AdS}_2$ solutions that we examine in this work is given by \cite{Lozano:2020txg}:
    \begin{equation}\label{eq: nsns_sector_full}
    \begin{aligned}
         \text{d}s^2 &=\frac{u}{\sqrt{h_4h_8}}\left(\frac{1}{4}\text{d}s^2_{\text{AdS}_2}+\frac{h_4h_8}{4h_4h_8+(u')^2}\text{d}s^2_{\text{S}^2}\right)+\sqrt{\frac{h_4}{h_8}}\text{d}s^2_{\text{T}^4}+\frac{\sqrt{h_4h_8}}{u}(\text{d}\rho^2+\text{d}\psi^2),\\
         e^{-2\Phi} &=h_8^2+\frac{h_8(u')^2}{4h_4},\\
         H_3 &=\frac{1}{2}\text{vol}_{\text{S}^2}\wedge\text{d}\left(\rho-\frac{uu'}{4h_4h_8+(u')^2}\right)+\frac{1}{2}\text{vol}_{\text{AdS}_2}\wedge\text{d}\psi,
    \end{aligned}
    \end{equation}
    where we wrote the above in the string frame. We have used $\Phi$ for the dilaton, $H=\text{d}B_2$ is the NS-NS flux 3-form and the functions $h_4$, $h_8$ and $u$ are functions of the $\rho$-coordinate\footnote{Note that in \cite{Lozano:2020txg} the authors have used both $h_4$ and $\hat{h}_4$ and in that context there is a subtle difference between the two. We will be using $h_4$ without hat here for convenience with the understanding that $h^{\texttt{here}}_4 = \hat{h}^{\texttt{there}}_4$.}\footnote{Note that there is another possibility where we have $h_4=h_4(\rho,z_{\text{CY}_2})$ and $h_8=h_8(\rho,z_{\text{CY}_2})$\cite{Lozano:2020txg}. We do not consider this kind of solutions in this paper.}. We have used $u'$ to represent $\partial_\rho u$, and the same is true for $h_4$ and $h_8$.  The $\psi$ is the T-dual-coordinate taking values within $[0,2\pi]$ as we have already explained. The torus is given by:
    \begin{equation}
        \text{d}s^2_{\text{T}^4} = \text{d}\theta_1^2+\text{d}\theta_2^2+\text{d}\theta_3^2+\text{d}\theta_4^2.
    \end{equation}
    
While it will not be necessary in our computations, for completeness we present the R-R sector of the theory. It is given by :
    \begin{equation}\label{eq: rr_sector_full}
    \begin{aligned}
        F_1&=h'_8\text{d}\psi,\\
        F_3&=-\frac{1}{2}\left(h_8-\frac{h'_8u'u}{4h_8h_4+(u')^2}\right)\text{vol}_{\text{S}^2}\wedge\text{d}\psi+\frac{1}{4}\left(\text{d}\left(\frac{u'u}{2h_4}\right)+2h_8\text{d}\rho\right)\wedge\text{vol}_{\text{AdS}_2},\\
        F_5&=-(1+\ast)h'_4\text{vol}_{\text{T}^4}\wedge\text{d}\psi=-h'_4\text{vol}_{\text{T}^4}\wedge\text{d}\psi+\frac{h'_4h_8u^2}{4h_4(4h_8h_4+(u')^2)}\text{vol}_{\text{AdS}_2}\wedge\text{vol}_{\text{S}^2}\wedge\text{d}\rho,\\
        F_7&=\frac{4h_4^2h_8-uu'h'_4+h_4(u')^2}{8h_4h_8+2(u')^2}\text{vol}_{\text{T}^4}\wedge\text{vol}_{\text{S}^2}\wedge\text{d}\psi-\frac{4h_4h_8^2-uu'h'_8+h_8(u')^2}{8h_8^2}\text{vol}_{\text{AdS}_2}\wedge\text{vol}_{\text{T}^4}\wedge\text{d}\rho,\\
        F_9&=-\frac{h_4h'_8u^2}{4h_8(4h_8h_4+(u')^2)}\text{vol}_{\text{AdS}_2}\wedge\text{vol}_{\text{T}^4}\wedge\text{vol}_{\text{S}^2}\wedge\text{d}\rho.
    \end{aligned}
    \end{equation}\par

It is obvious from \cref{eq: nsns_sector_full,eq: rr_sector_full} that the functions $u,h_4$ and $h_8$ determine completely the supergravity class of solutions. These functions are constrained such that the BPS equations and the Bianchi identities are satisfied. One can explicitly check that satisfying the type IIB equations of motion implies:
    \begin{equation}
     h''_4(\rho)=0, \qquad h''_8(\rho)=0
     \,             ,
    \end{equation}
    and
    \begin{equation}
        u''(\rho)=0.
    \end{equation}
    In the above, the first set of equations is a consequence of the Bianchi identities, while the condition on the function $u(\rho)$ is the BPS equation. In passing and for completeness, we mention that, in principle, we can have a violation of the Bianchi identities at some specific points where explicit brane sources are located. In this work we consider that the Bianchi identities always hold.
    
    Various cases for the definitions of $h_4$ and $h_8$ can be considered in relevance to the corresponding ones in the mother $\text{AdS}_3$ theories \cite{Lozano:2019zvg,Lozano:2019jza,Lozano:2019ywa}. In this work, we focus on an infinite class of solutions defined in a such a way that the these characteristic functions are piecewise continuous. In particular they are given by:
    \begin{equation}
    h_4(\rho)=\Upsilon
    \begin{cases}
    \beta_0\frac{\rho}{2\pi} &\quad 0\leq\rho\leq2\pi\\
    \sum\limits_{i=0}\limits^{k-1}\beta_i+\beta_k\left(\frac{\rho}{2\pi}-k\right) &\quad 2\pi k<\rho\leq2\pi(k+1),\quad k=1,\cdots,P-1\\
    \alpha_P+\alpha_P\left(-\frac{\rho}{2\pi}+P\right) &\quad 2\pi P<\rho\leq2\pi(P+1),
    \end{cases}
    \end{equation}
        \begin{equation}
    h_8(\rho)=
    \begin{cases}
    \nu_0\frac{\rho}{2\pi} &\quad 0\leq\rho\leq2\pi\\
    \sum\limits_{i=0}\limits^{k-1}\nu_i+\nu_k\left(\frac{\rho}{2\pi}-k\right) &\quad 2\pi k<\rho\leq2\pi(k+1),\quad k=1,\cdots,P-1\\
    \mu_P+\mu_P\left(-\frac{\rho}{2\pi}+P\right) &\quad 2\pi P<\rho\leq2\pi(P+1).
    \end{cases}
    \end{equation}
    In the above $\Upsilon$ is an overall normalisation\footnote{The precise relation is $\hat{h}_4 = \Upsilon h_4$, see e.g. \cite{Lozano:2019zvg}.}. The variables $\alpha_P,\beta_k,\mu_P,\nu_k$ with $(k=0,\cdots,P-1)$ are integration constants. Demanding that the functions $h_4(\rho)$ and $h_8(\rho)$ are continuous we have:
    \begin{equation}
        \alpha_P=\sum_{i=0}^{k-1}\beta_i,\qquad \mu_P=\sum_{i=0}^{k-1}\nu_i.
    \end{equation}
    
    We are interested in such a case that the $\rho$-coordinate defines a finite interval \cite{Lozano:2020txg} denoted by $\mathcal{I}_{\rho}$ within the range $[0,\rho^\ast]$. It is convenient to set the value $\rho^\ast$ to be equal to $\rho^\ast=2\pi(P+1)$, where $P$ is a large integer \cite{Lozano:2020txg}. In the class of solutions relevant to our studies, the functions $h_4(\rho)$ and $h_8(\rho)$ should vanish at the endpoints of the $\mathcal{I}_{\rho}$, namely $h_4(0)=h_8(0)=h_4(\rho^\ast)=h_4(\rho^\ast)=0$\footnote{We would like to stress that there are more exotic choices that could be made for the characteristic functions $h_4$ and $h_8$ related to the more exotic behaviour of the corresponding functions in the mother $\text{AdS}_3$ theories \cite{Filippas:2020qku}. We will not examine this kind of solutions here.}. We show an example of this behavior in \cref{fig: h4h8uplot}:
    
\begin{figure}[H]
\centering
\includegraphics[width=0.75\textwidth]{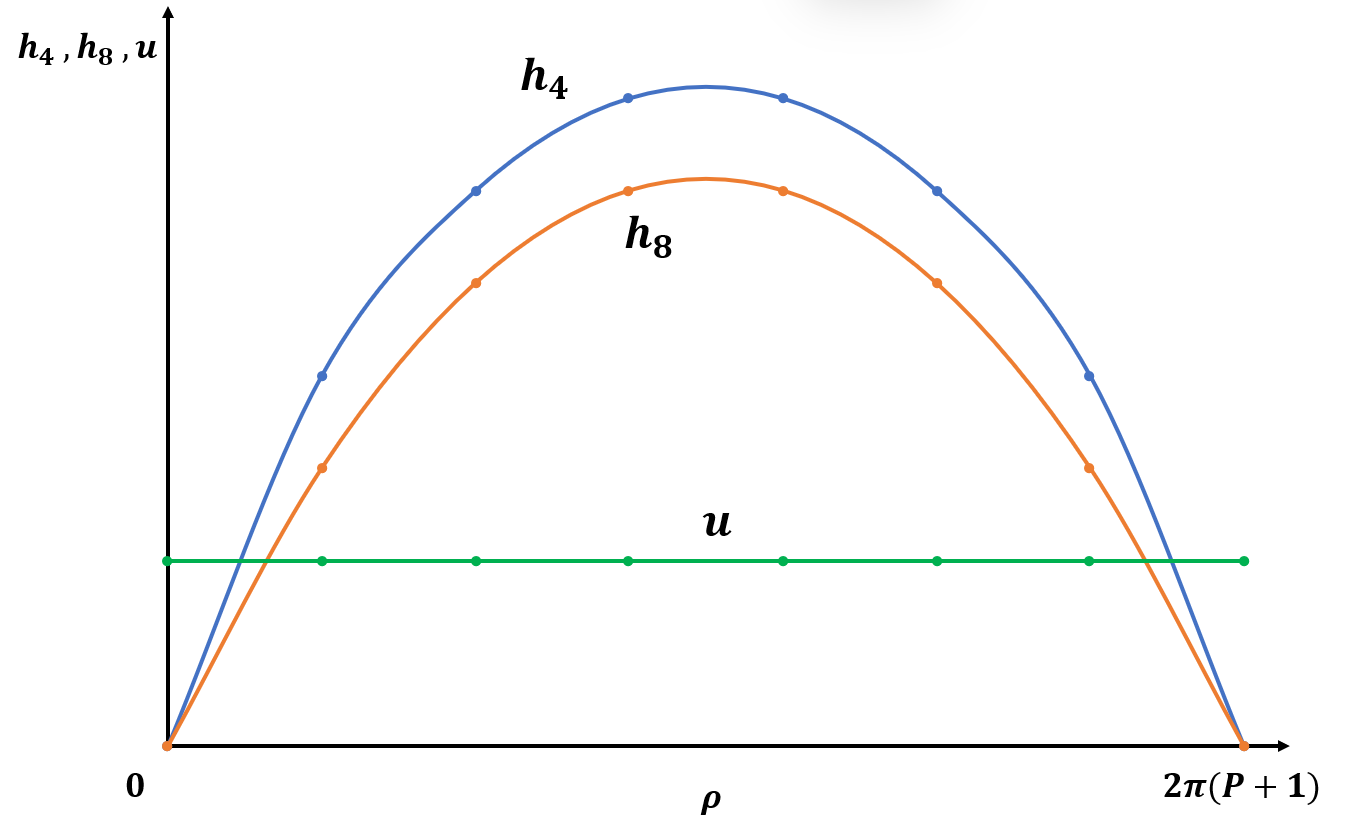}
\caption{An example of the characteristic functions $h_4(\rho), h_8(\rho)$ and $u$. Here we show a background for which the functions $h_4(\rho), h_8(\rho)$ vanish at the endpoints of the interval $\mathcal{I}_{\rho}$ and $u$ is being given by a constant.}
\label{fig: h4h8uplot}
\end{figure}
    
    It is clear from the BPS condition that the solution for the characteristic function $u(\rho)$ can be at most linear in $\rho$. Hence, we distinguish between two different behaviours that are equally well-defined:
    \begin{itemize}
        \item $u(\rho)=u_1\rho$,
        \item $u(\rho)=u_0$,
    \end{itemize}
    where in the above $u_{0,1}$ are just constants. With an eye towards the KK-spectrum, we mention that the case $u(\rho)=u_1\rho$ has been analyzed in \cite{Rigatos:2022ktp}, and thus in this work we focus on $u(\rho)=u_0$. To avoid cluttering our equations, we use the shortcut $u_0=u$ in the following. This result to the following simplification of the NS-NS sector:
    \begin{equation}\label{eq: metrics}
    \begin{aligned}
        &\text{d}s^2=\frac{u}{4\sqrt{h_4h_8}}(\text{d}s^2_{\text{AdS}_2}+\text{d}s^2_{\text{S}^2})+\sqrt{\frac{h_4}{h_8}}\text{d}s^2_{\text{T}^4}+\frac{\sqrt{h_4h_8}}{u}(\text{d}\rho^2+\text{d}\psi^2),\\
        &e^{-2\Phi}=h_8^2,\\
        &H_3=\frac{1}{2}(\text{vol}_{\text{S}^2}\wedge\text{d}\rho+\text{vol}_{\text{AdS}_2}\wedge\text{d}\psi),
    \end{aligned}
    \end{equation}
    and the relevant simplification of the R-R sectors readily follows, and we refrain from presenting the simplified  R-R sectors as it is not relevant in our computations. 
    
    The supergravity background presented above has the following brane realization, see \cref{branes living space}. 

    \begin{table}[H]
	\begin{center}
		\begin{tabular}{|c|c|c|c|c|c|c|c|c|c|c|c|}
			\hline	
 &&&&&&&&&&\\[-0.95em] 	    
			&	$x^0$	& $x^1$ 		& $x^2$ 		& $x^3$			& $x^4$ 		& $x^5$ 		& $x^6$ 		& 	$x^7$ 		& 	$x^8$ 		& 	$x^9$ 			\\ \hline \hline
	D1 		& 	--- 	& $\bullet$		& $\bullet$		& $\bullet$ 	& $\bullet$ 	& --- 			& $\bullet$  	&   $\bullet$	& 	$\bullet$  	&  $\bullet$ 		\\ \hline
	D3 		& 	---		& $\bullet$		& $\bullet$ 	& $\bullet$ 	& $\bullet$ 	& $\bullet$		& 	--- 		& 	--- 		& 	--- 		&  $\bullet$ 		\\ \hline
	D5 		& 	---		& --- 			& 	--- 		& --- 			& --- 			& --- 			& $\bullet$  	&   $\bullet$	&   $\bullet$	&  $\bullet$ 		\\ \hline
	D7 		& 	---		& ---			&	---  		& --- 			& --- 			& $\bullet$		& ---  			& 	--- 		& 	--- 		&  $\bullet$ 		\\ \hline
	NS5 	& 	---		& ---			&	---  		& --- 			& --- 			& $\bullet$		& $\bullet$  	&   $\bullet$	&   $\bullet$	& 	---  			\\ \hline
	F1 		&	---		& $\bullet$		& $\bullet$		& $\bullet$		& $\bullet$		& $\bullet$		& $\bullet$		& 	$\bullet$	& 	$\bullet$	& 	--- 			\\ \hline
		\end{tabular} 
\caption{A visualization of the supergravity solution given by \cref{eq: nsns_sector_full,eq: rr_sector_full} in terms of branes. We have used  (---) to denote that the brane extends along that direction. A $\bullet$ represents a coordinate that is transverse to the brane.}
\label{branes living space}
\end{center}
\end{table}
In \cref{branes living space}, $x^0$ is the direction associated with time in the ten-dimensional spacetime. $x^1, x^2, x^3, x^4$ are the directions of $\text{T}^4$. $x^5$ is the $\rho$-dimension, and $x^6, x^7, x^8$ are the three coordinates of the $\text{SO(3)}$ symmetry. Finally, $x^9$ is the $\psi$-internal.
\subsection{The field theory interpretation}
In this section we will discuss the basic features regarding the CFT which is dual to the AdS$_2$ background that we described in \cref{eq: nsns_sector_full,eq: rr_sector_full}. However, we find it more convenient to start by spelling out the basic characteristics of the $2$-dimensional seed CFT theories.  

\begin{figure}[H]
\centering
\includegraphics[width=0.75\textwidth]{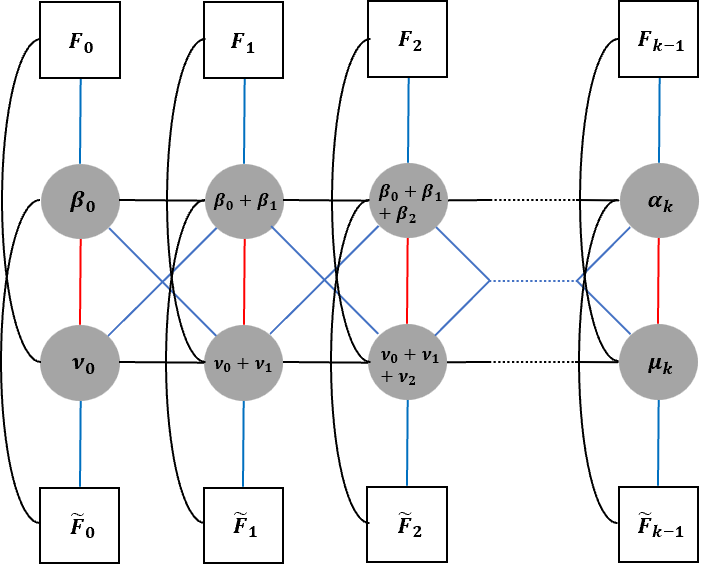}
\caption{The general 2d quiver field theory that is the seed of the CFT$_1$ considered in this work. The black line denotes a (4,4) hypermultiplet, the red line represents a (4,0) hypermultiplet and the blue line a (2,0) multiplet. Each gauge node has $\mathcal{N}=(4,4)$ vector multiplets degree of freedom.}
\label{fig: quiver plot}
\end{figure}

As we have already mentioned, the AdS$_2$ backgrounds that we examine in this work are related, by means of a T-duality transformation, to the seed AdS$_3$ solution of \cite{Lozano:2019jza}. In \cite{Lozano:2019zvg} it was shown that these AdS$_3$ solutions are the duals of the IR fixed-point limit of the quiver theory presented in \cref{fig: quiver plot}. In particular, the quiver diagram depicted in \cref{fig: quiver plot} flows to an IR fixed-point solution with dynamics that is captured by the AdS$_3$ background solution of \cite{Lozano:2019jza}. 

Let us explain the basic building blocks of the quiver diagram depicted in \cref{fig: quiver plot}. We have used a circle to denote a gauge node in the quiver with the associated group being $\text{SU}$. Hence, the $\beta_0$ in the circle means $\text{SU}(\beta_0)$ and likewise for the rest. We have, also, flavor groups which are denoted by squares and hence a square with $F_0$ is the flavor group $\text{SU}(F_0)$. The black lines connecting the different gauge nodes represent $\mathcal{N}=(4,4)$ hypermultiplets, and similarly, the red lines in here denote $\mathcal{N}=(4,0)$ hypermultiplets. We have used blue lines to denote $\mathcal{N}=(2,0)$ Fermi multiplets between two gauge nodes. Each gauge node is equipped with $\mathcal{N}=(4,4)$ vector multiplets degrees of freedom. 

Note that there are two rows of flavour groups, and the same is true for the color nodes. Crucially, the $F$'s and $\tilde{F}$'s are not independent of the other characteristic numbers of the quiver theory. The reasoning, as was explicitly observed and resolved in \cite{Lozano:2019zvg}, is that the field theory is chiral and hence we have to make sure to cancel gauge anomalies. This cancellation was carried out explicitly in \cite{Lozano:2019zvg}, leading to: 
\begin{equation}\label{eq: anomalycancellation}
    F_0 = \nu_0 - \nu_1
    \,      ,
    \qquad
    \tilde{F}_0 = \beta_0 - \beta_1
    \,      .
\end{equation}

We note that the above is an application to cancel out the anomalies at the first note and that we have to insist on anomaly cancellation at at each gauge node of the quiver. Similar relations can be derived straightforwardly, see \cite{Lozano:2019zvg} for further details. It is worthwhile to note that the numbers in \cref{eq: anomalycancellation} correspond to the number of color and flavor branes in the bulk description of the field theory. 

In \cite{Lozano:2020txg} the authors argued that the type IIB solutions they derived are dual to the IR limit of a $1$-dimensional quiver. This $1$-dimensional quiver inherits its shape directly from the seed $2$-dimensional theory that is depicted in \cref{fig: quiver plot}.

The way to think about the dynamics and matter content of the superconformal quantum mechanics theory is to regard it as the result of a dimensional reduction along the spatial direction of the $2$-dimensional $\mathcal{N}=(4,0)$ quivers that were previously derived in \cite{Lozano:2019zvg} and reviewed above. There are some additional twisted multiplets obtained after the dimensional reduction and we refer the reader to \cite[Appendix B]{Lozano:2020txg} for a detailed exposition.
\section{Holographic spin-2 fluctuations}\label{sec: spin2fluctuations}
    We start from the perturbations of the metric. We find it most convenient for the purposes of our analysis to re-write the metric in Einstein frame. To do so, we multiply by a factor of the dilaton, more precisely $e^{-\frac{\Phi}{2}}$. We will use the shorthand definitions for simplicity,
    \begin{equation}
    \begin{aligned}
        \hat{g}_{ab}\text{d}z^a\text{d}z^b&=e^{-\frac{\Phi}{2}}\left(\frac{u}{4\sqrt{h_4h_8}}\text{d}s^2_{\text{S}^2}+\sqrt{\frac{h_4}{h_8}}\text{d}s^2_{\text{T}^4}+\frac{\sqrt{h_4h_8}}{u}(\text{d}\rho^2+\text{d}\psi^2)\right)
        \,      ,
        \\
        f_1&=\frac{u}{4\sqrt{h_4h_8}}\,   ,
    \end{aligned}
    \end{equation}
    such that the metric in the Einstein frame becomes:
    \begin{equation}
        \text{d}s^2=e^{-\frac{\Phi}{2}}f_1\text{d}s^2_{\text{AdS}_2}+\hat{g}_{ab}\text{d}z^a\text{d}z^b.
    \end{equation}
    We consider fluctuations of the metric, which we will denote by $h$, along the directions of the $\text{AdS}_2$ part of the $10$-dimensional background. In particular:
    \begin{equation}
        \text{d}s^2=e^{-\frac{\Phi}{2}}f_1(\text{d}s^2_{\text{AdS}_2}+h_{\mu\nu}(x,z)\text{d}x^\mu\text{d}x^\nu)+\hat{g}_{ab}\text{d}z^a\text{d}z^b.
    \end{equation}
    We decompose $h_{\mu\nu}(x,z)$ into the transverse-traceless part in the bulk and a scalar in the internal part like:
    \begin{equation}
        h_{\mu\nu}(x,z)=h_{\mu\nu}^{(tt)}(x)\psi(z).
    \end{equation}
    Here the transverse-traceless tensor $h_{\mu\nu}^{(tt)}(x)$ satisfies:    
    \begin{equation}\label{eq: spin_2_ads_2_eigen}
        \Box_{\text{AdS}_2}^{(2)}h_{\mu\nu}^{(tt)}(x)=(M^2-2)h_{\mu\nu}^{(tt)}(x)=(\Box_{\text{AdS}_2}-2)h_{\mu\nu}^{(tt)}(x).
    \end{equation}
    The $\Box_{\text{AdS}_2}^{(2)}$ here is the Laplacian in $\text{AdS}_2$ acting on a rank-2 tensor and $\Box_{\text{AdS}_2}$ is the scalar Laplacian in $\text{AdS}_2$ \cite{Polishchuk:1999nh}.

    In order to make further progress, we recall that in \cite{Bachas:2011xa} the authors showed that the linearized Einstein equations can be reduced to the ten dimensional Laplace equation for the $h_{\mu \nu}$ given by:
    \begin{equation}\label{eq: laplacian_01}
        \frac{1}{\sqrt{-g}}\partial_M\sqrt{-g}g^{MN}\partial_Nh_{\mu\nu}=0.
    \end{equation}
    The above is the starting point of our analysis for the dynamics in the setup we consider here. In the above, the capital indices ${M,N}$ denote directions in the ten-dimensional background, and schematically are split into $M=\{\mu,a\}$; that is in an index taking values in the non-compact space, $\mu$, and another one along the compact directions. After separating \cref{eq: laplacian_01} into the two different parts as described above and a straightforward computation, we obtain: 
    \begin{equation}
        \hat{\Box}\psi(z)=-M^2\psi(z)        
        \,      ,
    \end{equation}
    where we have defined the operator $\hat{\Box}$ to be given by:
    \begin{equation}\label{eq: hatbox_def}
        \hat{\Box}\equiv\frac{1}{\sqrt{\hat{g}}}\partial_a(f_1e^{-\frac{\Phi}{2}}\sqrt{\hat{g}}\hat{g}^{ab}\partial_b),
    \end{equation}
    for notational convenience and we have also used \cref{eq: spin_2_ads_2_eigen} in deriving the above. As we see, by using the eigenvalues for the mode in the 2-dimensional AdS subspace, we have, now, simplified the 10-dimensional equations of motion to an 8-dimensional dynamical problem with a mass term. Upon evaluating the contributions from the different parts of the internal space, we arrive at:
    \begin{equation}\label{eq: eom_aux_1}
        \left(\nabla_{\text{S}^2}+\frac{u}{4h_4}\sum_{i=1}^n\partial_{\theta_i}^2+\frac{u^2}{4h_4h_8}(\partial_\psi^2+\partial_\rho^2)+M^2\right)\psi=0.
    \end{equation}
    
    In order to solve the differential equation obtained above we decompose the internal part of the metric perturbation $\psi$ into the appropriate eigenstates along the different parts of the internal space in the following manner:
    \begin{equation}\label{eq: mode_expansion}
        \psi=\sum_{\ell mnp}\Psi_{\ell mnp}Y_{\ell m}e^{in\cdot\theta}e^{ip\cdot\psi}
        \,      ,
    \end{equation}
    where in the above $n\cdot\theta=n_1\theta_1+n_2\theta_2+n_3\theta_3+n_4\theta_4$. We proceed by inserting the decomposition given by \cref{eq: mode_expansion} into \cref{eq: eom_aux_1} and we obtain: 
    \begin{equation}\label{eq: ODE}
        \frac{\text{d}^2}{\text{d}\rho^2}\Psi-\left(\frac{h_8n^2}{u}+p^2\right)\Psi+4\frac{M^2-\ell(\ell+1)}{u^2}h_4h_8\Psi=0
        \,      ,
    \end{equation}
    where in the above, to avoid cluttering our notation, we have suppressed the quantum numbers as subscripts; $\Psi_{\ell mnp}=\Psi$.
    
    This ordinary differential equation is a Sturm-Liouville problem. We can write it in a more suggestive way, namely in the standard way of expressing any Sturm-Liouville problem in the following way:
    \begin{equation}\label{sturm_liouville standard form}
        \frac{\text{d}}{\text{d}\rho}\left(L(\rho)\frac{\text{d}}{\text{d}\rho}\right)\Psi+Q(\rho)\Psi=-\lambda W(\rho)\Psi
        \,          ,
    \end{equation}
    with the characteristic quantities being given by: \begin{equation}\label{eq: sturm_liouville}
    \begin{aligned}
        L(\rho)&=1\,    , 
        \\
        Q(\rho)&=-\left(\frac{h_8n^2}{u}+p^2\right)\,   , 
        \\
        W(\rho)&=h_4h_8\,   , 
        \\
        \lambda&=4\frac{M^2-\ell(\ell+1)}{u^2}\,      ,
    \end{aligned}
    \end{equation}
and, of course, $\lambda$ being the eigenvalues of the Sturm-Liouville problem. The coordinate $\rho$ is valued in the range of $[0,2\pi(P+1)]$, where we remind the reader that $P$ is a large integer. We have already mentioned that the defining functions vanish at the endpoints of $\mathcal{I}_{\rho}$. This defines what is known in the mathematical literature as a singular Sturm-Liouville problem. We discuss the boundary conditions necessary to fully define the Sturm-Liouville problem in the next section, where we also proceed to derive its solutions.
\section{Unitarity bound and universal solutions}\label{sec: unitarityandsolutions}
    The next step we wish to make in this work is to obtain a bound for that mass eigenvalue $M^2$ from the Sturm-Liouville problem we derived in the previous section, see \cref{eq: ODE}. From this bound and by using the basic holographic relation between the mass and the conformal dimension, we will derive the unitarity bound. To obtain an inequality for $M^2$ from the equation of motion, we focus on a particular kind of solutions that is known in the existing literature as ``minimal universal class of solution''. Said class of solutions is universal in the sense that it does not depend on the exact form of $h_4$ and $h_8$.

    Let us begin by deriving the mass inequality that will eventually lead to the unitarity bound. To that end, we multiply the Sturm-Liouville problem, defined in \cref{eq: ODE}, by $\Psi$ and then integrate over the $\mathcal{I}_{\rho}$. This yields: 
    \begin{equation}
        \int_{\mathcal{I}_\rho}\Psi\frac{\text{d}^2}{\text{d}\rho^2}\Psi \text{d}\rho-\int_{\mathcal{I}_\rho}\left(\frac{h_8n^2}{u}+p^2\right)\Psi^2\text{d}\rho+4\frac{M^2-\ell(\ell+1)}{u^2}\int_{\mathcal{I}_\rho}h_4h_8\Psi^2\text{d}\rho=0,
    \end{equation}
    where we remind the reader that the interval $\mathcal{I}_{\rho}$ is bounded as $[0,\rho^\ast]$ and $\rho^\ast>0$. Firstly, we focus on the term that contains the second derivative of $\Psi$. We perform integration by part, such that we obtain:
    \begin{equation}\label{integration equation}
        4\frac{M^2-\ell(\ell+1)}{u^2}\int_0^{\rho^\ast}h_4h_8\Psi^2\text{d}\rho+\Psi\Psi'\big|_0^{\rho^\ast}=\int_0^{\rho^\ast}\left((\Psi')^2+\left(\frac{h_8n^2}{u}+p^2\right)\Psi^2\right)\text{d}\rho.
    \end{equation}
    In the above, $\Psi'$ is $\Psi$'s derivative of $\rho$.
    
    Although the interval $\mathcal{I}_{\rho}$ does not possess an exact U(1) isometry, we have seen that the $\rho^{\ast}$ endpoint is an integer multiple of $2\pi$. In other words, it is topologically, but not metrically, equivalent to $\text{S}^1$. Demanding single-valuedness for the wave-function leads to the choice of the boundary conditions $\Psi \big|_{\rho=0}=\Psi \big|_{\rho=\rho^{\ast}}$ and $\Psi' \big|_{\rho=0}=\Psi' \big|_{\rho=\rho^{\ast}}$. Focusing on the Hilbert space of such functions, it is clear that the term $\Psi \Psi'\big|_0^{\rho^\ast}$ vanishes. 
    
 Because $h_4,h_8,u,n,p$ are non-negative, we obtain straightforwardly:
    \begin{equation}\label{unitarity bound}
        M^2\geq \ell(\ell+1).
    \end{equation}

The next step is to obtain the ``minimal universal class of solutions''. We have already explained what ``universal'' is in this context. The ``minimal'' refers to this fact that we wish to saturate the bound we obtained earlier for $M^2$, see \cref{unitarity bound}. Upon setting $M^2 = \ell(\ell+1)$ the Sturm-Liouville problem given by \cref{sturm_liouville standard form,eq: sturm_liouville} reduces to:
    \begin{equation}\label{eq: mode_minimal}
    \frac{\text{d}^2}{\text{d}\rho^2}\Psi=0\,          ,
    \end{equation}
that has the obvious solution 
\begin{equation}\label{eq: mode_minimal_sltn}
    \Psi=a\rho+\Psi_0\,          ,
\end{equation}
where in the above $a$ and $\Psi_0$ are constants of integration. Notice, however, that we need to examine the compatibility of the general solution with the boundary conditions imposed earlier. A simple inspection yields: 
\begin{equation}\label{eq: mode_minimal_sltn_final}
    \Psi = \Psi_0\,          ,
\end{equation}
namely the ``minimal universal class of solutions'' that describe these spin-2 fluctuations is given by just a constant $\Psi_0$. Note that this situation is strikingly different compared to the solutions for a linear function $u(\rho)$ obtained in \cite{Rigatos:2022ktp}. Some commentary is in order. The modern intuition regarding these spin-2 operators in AdS$_2$ holography has been that the additional fluxes in the NS-NS and R-R sectors of the theory as well as the non-trivial warp factors in the geometry are sufficient to provide us with a richer spectrum in the CFT \cite{Lozano:2020txg,Lozano:2021rmk,Rigatos:2022ktp,Couzens:2021veb,Lozano:2022swp,Conti:2023naw}. Our findings suggest, however, that a more important role is played by the constraints arising from the BPS conditions of the background supergravity theory. Our reasoning for suggesting this comes from carefully observing the background for $u=\text{constant}$ given by \cref{eq: metrics}, for which we obviously have non-trivial fluxes and warp factors and the only different with \cite{Rigatos:2022ktp} is the solution to the BPS equations for $u$. 

The supergravity mode solutions that correspond to the metric perturbations we studied in this work are dual to operators in the field theory that carry conformal dimension $\Delta$. We can relate the mass, $M^2$, of the supergravity states to the conformal dimension of the operators using the standard AdS/CFT formula
\begin{equation}
    M^2=\Delta(\Delta-1).
\end{equation}
From the inequality we derived earlier, see \cref{unitarity bound}, we can conclude for the scaling weight of the operators a lower bound given by
\begin{equation}
    \Delta\geq\Delta_{\text{min}}, \qquad\text{with}\qquad  \Delta_{\text{min}}=\ell+1.
\end{equation}
We note that we can, in principle, study non-minimal solution, namely solutions for which either $M^2>\ell(\ell+1)$ or there are non-trivial excitations along T$^4$ or $\mathcal{I}_\rho$. However, in order to study this latter kind of solutions we need to make very specific choices for $h_4$ and $h_8$ and we do not examine these solutions here.
\section{The holographic central charge}\label{sec: centralcharge}

The analysis we performed in the previous sections has provided us with all the necessary information, in order to proceed to the computation of the central charge for the class of solutions described in \cref{eq: metrics}. This is the purpose of this section.

Before we proceed with the computational part of this section, however, we feel that some comments are in order. Defining the notion of the central charge in a one-dimensional conformal field theory is a delicate and subtle issue. The reasoning is quite straightforward. Let us consider a general one-dimensional conformal field theory. The energy-momentum tensor, denoted by $T_{\mu \nu}$, has only one component. Let us denote this component by $T_{00}$. Then by virtue of the tracelessness property, this component vanishes, $T_{00}=0$.

Taking the above into consideration, we wish to clarify that the quantity we call the ``holographic central charge'' has the interpretation as the number of vacuum states in the dual conformal quantum mechanics theory. Our reasoning for choosing to name said quantity as the central charge is to remain consistent with past literature.

In order to be concrete on how we make progress with this computation, we borrow intuition from higher-dimensional examples where this methodology has been successfully applied see, e.g. \cite{Gutperle:2018wuk,Chen:2019ydk,Speziali:2019uzn}. In these examples, using the unitarity bound and the minimal universal class of solutions, the analogous expressions to the ones we have obtained in the previous sections, the task at hand was to compute the normalization of the two-point functions of the operators $T_{\mu \nu}$. This is the operator that is sourced by the the graviton fluctuations. For the case of the minimal solution with no excitation in the internal space the supergravity mode is that of the massless graviton fluctuation, which is dual to the  the energy-momentum tensor. Using holography, the normalization of the  two-point function of the energy-momentum tensor
can be determined from the effective action for the lower-dimensional graviton\footnote{By lower-dimensional graviton we mean that in an $\text{AdS}_5$ theory we have the five-dimensional graviton and likewise for other examples.}. While some might argue that this is a flawed analogy, recent studies suggest the same principles also apply to the $\text{AdS}_2$ case since the computation of the holographic central charge agrees with the entanglement entropy as they should \cite{Lozano:2020txg,Rigatos:2022ktp}. This gives us confidence on how to proceed and indeed we will see that even when the function $u$ is simply given by a constant the final result is consistent with the expression of the entanglement entropy.

Our starting point is the Einstein-frame Type IIB action which we can write schematically in the following form:
\begin{equation}\label{eq: typeiib_action}
    S_{\text{IIB}}=\frac{1}{2\kappa_{10}^2}\int\sqrt{-g}(R+\cdots)\text{d}^{10}x
\end{equation}
Note that in the above $\cdots$ represent terms in the action that are irrelevant to our computation.

Expanding the action given by \cref{eq: typeiib_action} to quadratic order in the metric fluctuations yields:
\begin{equation}\label{eq: action_centrlacharge}
\delta^2S=\frac{1}{\kappa_{10}^2}\int h^{\mu\nu}\partial_M\sqrt{-g}g^{MN}\partial_Nh_{\mu\nu}\text{d}^{10}x+\ldots
\end{equation}
where in the above $\ldots$ denote a boundary term. Expanding out the action given by \cref{eq: action_centrlacharge} explicitly and dropping the boundary term leads to:
\begin{equation}
    \delta^2S[h_{\mu\nu}]=\frac{1}{\kappa_{10}^2}\int\sqrt{-g_{\text{AdS}_2}}\sqrt{\hat{g}}h^{\mu\nu}\left(\Box_{\text{AdS}_2}^{(2)}+2+\hat{\Box}\right) h_{\mu\nu}\text{d}^{10}x
\end{equation}
We remind the reader that the operator $\hat{\Box}$ in the above relation is defined in \cref{eq: hatbox_def}.

Expanding the $h_{\mu \nu}$ mode according to \cref{eq: mode_expansion}
\begin{equation}
    h_{\mu\nu}=\sum_{\ell mnp}(h_{\ell mnp}^{(tt)})_{\mu\nu}Y_{\ell m}\Psi_{\ell mnp}e^{in\cdot\theta}e^{ip\cdot\psi}
\end{equation}
leads to
\begin{equation}
    \delta^2S=\sum_{\ell mnp}C_{\ell mnp}\int_{\text{AdS}_2}\sqrt{-g_{\text{AdS}_2}}(h_{\ell mnp}^{(tt)})^{\mu\nu}\left(\Box_{\text{AdS}_2}^{(2)}+2-\ell(\ell+1)\right)(h_{\ell mnp}^{(tt)})_{\mu\nu}\text{d}^2x
\end{equation}
with spherical harmonics begin normalized as:
\begin{equation}
\int Y_{\ell m}Y_{\ell'm'}\text{d}^2x=1
\end{equation}
and the coefficients $C_{lmnp}$ being given by:
\begin{equation}\label{eq: clmnp_def}
C_{\ell mnp} = \frac{1}{\pi} \int_{\mathcal{I}_{\rho}} \sqrt{\hat{g}} (e^{-\tfrac{\Phi}{2}} f_1)^{\tfrac{1}{2}} |\Psi_{\ell mnp}|^2
\end{equation}

The integral in \cref{eq: clmnp_def} is finite for the fluctuations we have considered in this work, since they remain finite at every point of the background spacetime. We specialize to the ``minimal universal solution'' obtained in \cref{sec: unitarityandsolutions}. Unlike the situation in \cite{Lozano:2020txg}, in our case $\Psi$ is just a constant. Setting all the quantum numbers to be equal to zero, we can set $|\Psi_{0000}|^2=1$, such that we obtain: 
\begin{equation}\label{central charge}
    C_{0000}=\frac{1}{2\pi}\int_0^{\rho^\ast}h_4h_8\text{d}\rho,
\end{equation}
where in deriving the above we have used the standard results:
\begin{equation}
     \text{vol}_{\text{S}^2}=4\pi,
     \quad 
     \text{vol}_{\text{T}^4}=(2\pi)^4, 
     \quad
     \text{vol}_{\mathcal{I}_\psi}=2\pi,
     \quad
     2\kappa_{10}^2=(2\pi)^7.
\end{equation}

The above result, \cref{central charge}, agrees with the computation of holographic entanglement entropy derived in \cite{Lozano:2020txg}, up to an irrelevant numerical factor, as it should. 
\section{Discussion and outlook}\label{sec: outlook}
    In this paper we considered aspects of spin-2 perturbations around a warped solution in the Type IIB theory of the schematic form $\text{AdS}_2 \times \text{S}^2 \times \text{T}^4 \times \mathcal{I}_\psi \times \mathcal{I}_\rho$. The solution comes equipped with fluxes in the NS-NS and R-R sectors. They are derived after a T-duality transformation from a seed $\text{AdS}_3 \times \text{S}^2 \times \text{T}^4 \times \mathcal{I}_\rho$ family of solutions in massive Type IIA supergravity. We have derived the equations of motion governing the dynamics of these spin-2 states following the blueprint of \cite{Bachas:2011xa}. As expected from earlier studies, we can distinguish them into two broad classes; the universal and non-universal. We note that the universal class of solutions are independent of the defining functions of the background, and thus are present in all 1-dimensional field theories dual to the backgrounds examined here. 
    
    In this work we have explicitly computed the so-called ``minimal universal class of solutions''. This is a class of solutions that saturates the bound on the mass. These supergravity states source operators in the dual field theory of classical dimension $\Delta=\ell+1$, where $\ell$ is the quantum number on the $\text{S}^2$; the holographic realization of the $SU(2)_R$ R-symmetry. 
    
    Using this class of solutions and the expansion of supergravity action we were able to compute the expression for the holographic central charge. We checked that it agrees with the computation of the holographic entanglement entropy that was performed in \cite{Lozano:2020txg} as it should.
    
    We mention that our work is complementary to \cite{Rigatos:2022ktp}. The difference in these two works is the choice of the function $u(\rho)$. In particular, we chose it to be a constant, while \cite{Rigatos:2022ktp} examined a linear solution. The difference between the different choices of $u(\rho)$ is reflected on the solutions of the metric perturbations $h_{\mu\nu}(x,z)$. More specifically, we found that $u(\rho)=\text{constant}$ leads to a solution given by a constant, while the non-trivial functional dependence of $h_{\mu\nu}(x,z)$ is achieved only for $u(\rho)$ being a linear function. Of course the expressions for the unitarity bound and central charge agree in the two pictures.
    
    Our results lead us to the following interpretation. In backgrounds with pure $\text{AdS}_2$ factors, it was well-understood that the modes examined here are non-trivial if and only if the mass is zero, see e.g \cite{Michelson:1999kn,Maldacena:1998uz,Lee:1999yu,Corley:1999uz}. The more recent explorations of $\text{AdS}_2$ supergravity solutions have been suggesting that non-trivial warp factors and fluxes lead to a richer and more exotic description \cite{Lozano:2020txg,Rigatos:2022ktp,Conti:2023naw}. However, our choice of $u$ led to trivial spin-2 states without making the warp factors and fluxes equal to some constants. Therefore, we conclude that a more pivotal role is played by the supersymmetry conditions.
    
    Before we finish this section we would like to discuss some interesting future topics. 
    
    Based on \cite{Speziali:2019uzn,Rigatos:2022ktp}, we know that we can extract concrete information for the seed $\text{AdS}_3$ theories from the daughter $\text{AdS}_2$ and vice versa. More precisely, it was found that the expressions for the ``minimal universal class of solutions'' and central charge of the theories were the same. It would be interesting to examine if this connections persists for the choice $u=\text{constant}$ by a direct computation. This would be strong indicative suggestions on the importance of the BPS conditions in this type of quiver theories.
    
    Another interesting avenue of future work is to delineate more aspects of the dual superconformal quantum mechanics following the recent progress of \cite{Filippas:2020qku} in the mother 2-dimensional theories. In that work the author found that in some cases the dual quiver theories were anomalous and additional matter had to be added, in order to cancel gauge anomalies. The way that this mechanism was realized in supergravity was via bound states of D-branes. Interestingly, these anomalous quiver arose for choices of both $u(\rho)$ given by a linear function or a constant. Owing to the connections of the 2-dimensional CFT to the conformal quantum mechanics, it would be very desirable to understand these phenomena in the $\text{AdS}_2$ solutions.
    
    Furthermore, another interesting question is the status of (non)integrability and chaos in these theories. This is a rich subject, with well-developed analytical and numerical methods. We cannot do justice to the literature, so we mention as representative examples \cite{Basu:2011di,Basu:2011fw,Filippas:2019ihy,Filippas:2019bht,Rigatos:2020igd,Nunez:2018ags,Nunez:2018qcj,Rigatos:2020hlq,Pal:2023ckk}. The gist of the approach is to carefully consider the embedding of a fundamental string in the background and examine the resulting dynamics.

    Finally, there has been recent activity in the construction of warped $\text{AdS}_2$ solutions that have a holographic realization as defect within higher-dimensional solutions \cite{Lozano:2020sae,Lozano:2021rmk,Itsios:2023uae,Lozano:2022swp,Lozano:2021fkk}. We find it interesting to address the question of the spectrum in those theories as well.
    
    We hope to report on some of the above in the near future.
    
\newpage
\section*{Acknowledgments} 
We thank Carlos Nunez and Xinan Zhou for feedback on the draft and Kostas Rigatos for suggesting this problem to us. We are, also, grateful to Xinan Zhou and Kostas Rigatos for help with the English language in writing this paper. The work of Shuo Zhang is supported by funds from the Kavli Institute for Theoretical Sciences (KITS).
\newpage
\bibliographystyle{ssg}
\bibliography{spin2bib}
\end{document}